%Paper: hep-th/9205028
%From: Andy Strominger <andy@denali.physics.ucsb.edu>
%Date: Tue, 12 May 92 18:09:28 PDT

%The jnl.tex macros necessary to tex this are available from hepth.
%There are two figures appended at the end of this file.

\input jnl
\preprintno{UCSBTH-92-18}
\preprintno{hepth@xxx/9205028}
\def\DDD{I\kern -0.35em D\ }
\def\RRR{I\kern -0.35em R\ }
\def\nnn{/\kern -0.60em \nabla\ }

\input reforder
\ignoreuncited

\title            FADEEV-POPOV GHOSTS AND ${1} + {1}$
                  DIMENSIONAL BLACK HOLE EVAPORATION
\bigskip
\author           ANDREW STROMINGER

\affil \ucsb      %IF THE AUTHOR IS NOT FROM UCSB DELETE \ucsb AND ENTER
	          %THE AUTHOR'S ADDRESS PUTTING EACH LINE OF THE ADDRESS
	          % ON A DIFFERENT LINE OF YOUR FILE.

andy@denali.physics.ucsb.edu

\abstract
Recently Callan, Giddings, Harvey and the author derived a set of one-loop
semiclassical equations describing black hole formation/evaporation in
two-dimensional dilaton gravity conformally coupled to ${N}$ scalar fields.
These equations were subsequently used to show that an incoming matter wave
develops a black hole type singularity at a critical value ${\phi_{cr}}$ of
the dilaton field. In this paper a modification to these equations arising
from the Fadeev-Popov determinant is considered and shown to have dramatic
effects for ${N}<{24}$, in which case ${\phi_{cr}}$ becomes complex.  The
${N}<{24}$ equations are solved along the leading edge of an incoming
matter shock wave and found to be non-singular. The shock wave arrives at
future null infinity in a zero energy state, gravitationally cloaked by
negative energy Hawking radiation. Static black hole solutions supported by
a radiation bath are also studied. The interior of the event horizon is
found to be non-singular and asymptotic to deSitter space for ${N}<{24}$,
at least for sufficiently small mass. It is noted that the one-loop
approximation is {\it not} justified by a small parameter for small ${N}$.
However an alternate theory (with different matter content) is found for
which the same equations arise to leading order in an adjustable small
parameter.
\endtitlepage
In recent work [\cite{Evan}] it was proposed that two-dimensional dilaton
gravity coupled to conformal matter is a useful and simple model in
which progress might be made in unravelling the mysteries associated
with black hole evaporation [\cite{Hawk}]. It was shown that the
process of black hole formation/evaporation, in an approximation which
includes one-loop matter effects and treats gravity semiclassically,
can be described by a set of partial differential equations which incorporate
the backreaction of Hawking radiation on the geometry. It was further
pointed out that this approximation is formally exact in a limit as
the number ${N}$ of matter fields is taken to infinity.

It was subsequently shown [\cite{rst,ban}] that in the large ${N}$
approximation a collapsing matter wave still forms a black hole containing a
singularity\footnote{*}{Or more precisely, the fields become so large
that the large-${N}$ approximation breaks down.}.
This singularity no longer occurs at the value
${\phi} = {\infty}$ of the dilaton
as in the classical theory, but rather moves up to the finite
value ${\phi_{cr}} = {-} {{1}\over{2}}\,{\ell}{n}\,{{N}\over{12}}$. The
black hole then evaporates, presumably leaving a massless, singular
``remnant'' [\cite{sth,hwk,qbh}].

In this paper we will consider the equations describing dilaton gravity
coupled to ${N}$ conformal scalars in the one-loop approximation for
finite ${N}$. These equations differ from those derived in [\cite{Evan}]
by the addition of terms arising from the gravity-ghost measure which are
negligible for ${N}{\rightarrow}{\infty}$. We shall see that for
${N}<{2}{4}$, these terms remove the singularity found in
[\cite{rst,ban}].

Some evidence consistent with the absence of other types of singularities
is presented, but the equations are sufficiently complex that the
question is not settled here. We hope to analyze the problem numerically
in the near future [\cite{bist}].

We begin with a discussion of the gauge fixing and quantization of pure
dilaton gravity
$$
{S_0} = {{1}\over{2\pi}}{\int}{d^2}{\sigma}{\sqrt{-g}}{e^{-2\phi}} [{R}
+{4} ({\nabla}{\phi})^2 + {4}{\lambda^2}],\eqno(dlct)
$$
\noindent where ${g}$ and ${\phi}$ are the metric and dilaton fields
respectively
and ${\lambda^2}$ is a cosmological constant.\footnote{*}{A related discussion
of this quantization with similar conclusions has been given in
[\cite{hver}].}
This is a theory with no local degrees of freedom: the 3 + 1 fields in
${g}$ and ${\phi}$ may be eliminated by two gauge conditions and two
constraints. There are however a one parameter family of classical black
hole solutions labelled by the black hole mass [\cite{Witt}].
We wish to
gauge fix (\call{dlct}) to conformal gauge
$$\eqalignno{
{g_+}\,{_-}&= {-}{{1}\over{2}}\,{e^{2\rho}},\cr
{g_{++}}&={g_{--}}={0},&({gfcr})\cr}
$$
\noindent where ${\sigma}^{\pm} = {\tau}{\pm}{\sigma}$. In so doing the
action will be shifted  by the usual logarithm of the Fadeev--Popov ghost
determinant. This term may be expressed in a covariant notation as
$$
{S_{FP}} = {{13}\over{{4}{8}{\pi}}}{\int}{d^2}{\sigma}{\sqrt{-g}}
{R}{\square^{-1}}{R}.\eqno(fpct)
$$
\noindent However an ambiguity in this procedure arises in the present
context. There is a family of metrics ${g_{\alpha}}$ given by
$$
{g_{\alpha}} = {e^{-2\alpha\phi}}{g}\eqno(glph)
$$
\noindent any of which might be used to construct ${S_{FP}}$ in
(\call{fpct}). Although we have
chosen ${g_0}$ in order to write down (\call{dlct}), it is not especially
preferred. Indeed the difference between two choices of metric is given
by
$$
{S_{FP}}({g_{\alpha}}) - {S_{FP}}({g}) = {{13\alpha}\over{12\pi}}
{\int}{d^2}{\sigma}{\sqrt{-g}} ({R} + {\alpha}{\square}{\phi})
{\phi}.\eqno(delfp)
$$
\noindent This is a local expression which might have been added to the
action (\call{dlct})
either in the first place or as a finite counterterm during
one-loop renormalization.
Thus there is no right or
wrong choice of metric in (\call{fpct}): different choices simply correspond
to different theories. One must choose a theory which contains the physical
phenomena one wishes to investigate.

In fact ${g}$ is not a good choice of metric to use for defining the ghost
measures--it leads to a sick theory. Using the metric ${g}$ in
(\call{fpct}) means that the Fadeev-Popov ${b-c}$ ghosts couple to the geometry
in the same way as the conformal ${f}$ fields described in
[\cite{Evan}]. It immediately follows that black holes will grow in mass by
Hawking radiation of negative energy ghosts.\footnote{*}
{Although we cannot build
detectors to see the ghosts directly, we can still observe their effects
on the geometry.} This is clearly nonsense.

This problem is avoided by defining the ghost measure with the alternate
metric
$$
{\hat g} = {e^{-2\phi}}{g}.\eqno(ghat)
$$
\noindent This metric turns out to be flat for all classical solutions
of (\call{dlct}). Black holes will therefore not radiate ghosts to leading
order.

As is familiar in Liouville gravity, there is an additional term of the
form (\call{fpct}) arising from the dependence of the ${\rho}, {\phi}$
measures on the metric.   Again there is an ambiguity in these measures.
Since ${\rho}, {\phi}$ are not local, propagating degrees of freedom, it is
natural to demand that there be no Hawking radiation in these modes.
This is accomplished by using the metric ${\hat g}$ to define their
measures as well, which changes the 13 in equation (\call{fpct})
to a 12. This definition
ensures that there is a stable black hole
solution of the quantum theory for each value of the mass ${M}$.

The gauge fixed action, including all the measure terms, is then
$$\eqalignno{
{S_0}{+}{S_M}&= {{1}\over{\pi}}{\int}{d^2}{\sigma} [({e^{-2\phi}}(2{\partial_+}
{\partial_-}{\rho} - {4}{\partial_+} {\phi}{\partial_-}{\phi} {+}
{\lambda^2}{e^{2\rho}})\cr
&\qquad {+}{2}{\partial_+}({\rho} - {\phi}) {\partial_-}({\rho} - {\phi}))],
&({rhct})\cr}
$$
\noindent while the stress tensor is\footnote{**}
{Ignoring the cosmological constant term, (\call{rhct}) becomes
a free theory in terms of the variables ${v} = {e^{-2\phi}}$ and
${w} = {\rho} - {\phi}$. It is then easy to check that the stress tensor has
${c} = {26}$, as required by coordinate invariance. It also easily
follows that the cosmological constant operator ${e^w}$
is dimension  (1,1) (with no renormalization of the exponent). Thus there are
many similarities with Liouville
theory, and the methods developed there may be useful in the present context.}
$$
{T_{++}^0} {+}
{T_{++}^M} = {e^{-{2}{\phi}}}({4}{\partial_+}{\phi}{\partial_+}{\rho}{-}{2}
{\partial_+^2}{\phi}) +
{2}({\partial_+}{\rho} - {\partial_+}{\phi})^2 - {2} {\partial_+^2}
{\rho} {+} {2}{\partial_+^2}{\phi} + {t_+}.\eqno(rht)
$$
\noindent where, as explained in [\cite{Evan}], ${t_+}({\sigma^+})$ is
determined
by boundary conditions.

\indent In order to study the problem of black hole evaporation we must
complicate
the theory by adding matter fields with local degrees of freedom. Following
[\cite{Evan}] we add ${N}$ conformally coupled scalar matter fields ${f_i}$.
Again an ambiguity in defining the ${f}$ measure arises. However this
time we do not wish to use the metric ${\hat g}$. This leads to
a presumably sensible theory which does not contain the phenomena we
wish to study: black holes do not Hawking radiate. Indeed this theory in
a sense does not even contain black holes, since the matter sees only the
flat metric ${\hat g}$. As explained in [\cite{Evan}], if we instead use ${g}$
to define the ${f}$ measure Hawking radiation of ${f}$ particles indeed
occurs, and closely resembles the four-dimensional phenomena.
One thereby arrives at the final action
$$\eqalignno{
{S}&={{1}\over{\pi}}{\int}{d^2}{\sigma}
[{e^{-2\phi}} ({2}{\partial_+}{\partial_-}{\rho} - {4}{\partial_+}{\phi}
{\partial_-}{\phi}
+{\lambda^2}{e^{2\rho}})\cr
&\qquad - {{N}\over{12}}\,\,{\partial_+}{\rho}{\partial_-}
{\rho}
+{{1}\over{2}}\,\,{\sum_{i=1}^{N}}{\partial_+}{f_i}{\partial_-}{f_i}
{+}{2}{\partial_+}({\rho} - {\phi}){\partial_-}({\rho} - {\phi})].
&({ghct})\cr}
$$
\noindent The constraints will be discussed
shortly.

The quantum theory is described by functional integration with
the ``naive'' measure weighted by ${S}$. In the large-${N}$ limit, all terms
are of order ${N}$ (after shifting ${\phi}$)
except the last one, which is order one and may therefore
be dropped. In [\cite{Evan}] it was argued that these order ${N}$ terms may be
treated as a quantum effective action which describes the process of black
hole formation and evaporation, with the
modifications of the gravitational
action accounting for the stress-energy carried by Hawking radiation.

However, the large-$N$ approximation is not necessarily a reliable barometer of
finite-$N$ physics,
particularly for $N \le 24$. One
way to see this is from the behavior of the $\rho - \phi$
kinetic operator $\cal K$ at large positive and negative
$\phi$. At large negative $\phi$, the theory is essentially classical and $\cal
K$
has one positive and one negative eigenvalue. At large positive $\phi$ and
finite $N$, the
classical action may be treated as a perturbation about the free
measure-induced terms, and
$\cal K$ still has one positive and one negative
eigenvalue\footnote{*}{Although for $N>24$ there is
a region whose size grows with $N$ in which there are two negative
eigenvalues.}.
In the large-$N$ limit however, there are two negative eigenvalues for large
$\phi$ and
consequently a zero eigenvalue at an intermediate value of $\phi$. This
zero eigenvalue leads to singular behavior in the large-$N$
limit[\cite{rst,ban}] which may not be present for
small $N$. Thus other methods should be found for analyzing the theory at small
$N$.

In this paper the action
(\call{ghct}) including the last term (and corresponding
modification of the constraints) will be treated as a quantum effective
action for finite ${N}$. This amounts to a one-loop semiclassical
approximation. For small ${N}$, there is no obvious small parameter which
justifies this approximation. The loop expansion may break down
when ${e^{2\phi}}$ gets large, and we cannot be confident that our conclusions
are
qualitatively correct. Nevertheless, we shall find in the one-loop
approximation that the behavior of the theory changes dramatically at
${N} = {24}$, and we hope that the one-loop semiclassical approximation
is at least qualitatively correct in the $N<24$ regime.

While treating (\call{ghct}) semiclassically has not been justified
as a systematic approximation to dilaton gravity coupled to ${N}$
scalar fields for small ${N}$, it can be formally justified as a systematic
approximation to dilaton gravity coupled to a different matter
system: let there be ${NM}$ scalar ${f}$ fields, and include an
additional ${c} = {-}{24}{M}$ conformal matter sector with measure
defined with respect to ${\hat g}$. After a shift of ${\phi}$, one
recovers an action of the form (\call{ghct}) multiplied by ${M}$. One
then expects a semiclassical treatment to be valid for large ${M}$.
The following analysis may be taken to apply to this system.

We now proceed to analyze the dynamics following from (\call{ghct}).
The ${\rho}$ and ${\phi}$ equations may be cast in the useful form
$$\eqalignno{
{8}{P}{\partial_+}{\partial_-}{\phi}&= {-} {P^{\prime}} ({4}{\partial_+}
{\phi}{\partial_-}{\phi} + {\lambda^2}{e^{2\rho}}),&({pheq})\cr
{2}{P}{\partial_+}{\partial_-}{\rho}&={e^{-4\phi}}
({4}{\partial_+}{\phi}{\partial_-}{\phi}+{\lambda^2}{e^{2\rho}}),
&({rheq})\cr}
$$
\noindent where
$$\eqalignno{
{P}\,\,&{\equiv}\,\,
{e^{-4\phi}}-{{N}\over{12}}\,\,{e}^{-2\phi}+{{N}\over{24}},\cr
{P^{\prime}}\,\,&{\equiv}\,\,{{{\delta}{P}}\over{{\delta}{\phi}}} =
{4}
{e^{-2\phi}}({{N}\over{24}} - {e^{-2\phi}}).&({ppeq})\cr}
$$
\noindent The $++$ constraint equation is
$$\eqalignno{
{T_{++}}&={e^{-2\phi}}({4}{\partial_+}{\phi}{\partial_+}
{\rho} - {2}{\partial_+^2}{\phi})
{+}{{1}\over{2}}\,\,{\sum_{i=1}^{N}}{\partial_+}{f_i}{\partial_+}
{f_i}\cr
&\qquad {-} {{N}\over{12}} ({\partial_+}{\rho}{\partial_+}{\rho} -
{\partial_+^2}{\rho})
{+} {2} ({\partial_+}({\rho}-{\phi}){\partial_+}({\rho}-{\phi})
- {\partial_+^2}({\rho}-{\phi})) + {t_+} = {0},&({+ppl})\cr}
$$
and a similar equation holds for ${T_{--}}$.

The effect of the ghost induced-terms in these equations is immediately evident
from (\call{pheq}). The prefactor ${P}$ in (\call{pheq}) has zeros at
$$
{e^{-2\phi}} = {{N}\over{24}}\,\,
({1}{\pm}{\sqrt{{1}-{{24}\over{N}}}}\,).\eqno(zros)
$$
\noindent As pointed out in [\cite{rst,ban}], these zeros are very
dangerous: because the RHS of (\call{pheq}) is generically non-zero,
${\partial_+}{\partial_-}{\phi}$ is forced to diverge whenever ${\phi}$
crosses a zero.

However for ${N}<{24}$ there are no real solutions of (\call{zros}),
and ${P}$ is a positive definite quantity with a minimum at ${\phi}
= {-} {{1}\over{2}}{\ell}{n}\,\,{{N}\over{24}}{~~}{\equiv}{~~}{\phi_c}$.
\footnote{${\dagger}$}{The stability of the zeros (or lack thereof)
of ${P}$ -- on which our results strongly depend -- against higher
loop quantum corrections is an important question to which we do not
have a definitive answer. However it can be easily seen,
by considering perturbation theory in ${e^{\phi}}$ and ${e}^{-\phi}$,
that in the large $N$ limit (in which the last term in $P$ is
neglected), ${P}$ must change sign between weak and strong
coupling, and consequently must have at least one zero. For ${N}<{24}$,
it does not change sign, and must therefore have an even number of zeros,
though we are not sure if that even number is zero in the exact theory.}

Thus the singularities described in [\cite{rst,ban}] do not arise.
While we do not know
if other types of singularities arise in the ${N}<{24}$ equations, they
are clearly far better behaved.\footnote{*}{The weak-coupling singularities
of the quantum kink solutions found in [\cite{sth,qbh}], are
presumably still present.}

These equations can be solved following [\cite{rst,ban}] perturbatively
about the leading edge of an ${f}$ shock wave incident on the linear
dilaton vacuum, as illustrated in Figure 1. The linear dilaton vacuum is a
solution of
(\call{pheq})--(\call{+ppl}) given by
$$\eqalignno{
{\phi}&={-}{\lambda}{\sigma},\cr
{\rho}&={0}.&({ldvm})\cr}
$$
\noindent An ${f}$ shock wave is defined by
$$
{{1}\over{2}}\,\,{\sum_{i=1}^{N}}{\partial_+}{f_i}{\partial_+}{f_i}
\,\,{\equiv}\,\,
{T_{++}^{f}} = {M}{\delta}({\sigma^+}
- {\sigma_0^+}).\eqno(fshw)
$$
\noindent Classically, the ${f}$ stress-energy leads to a black hole
of mass ${M}$ above
${\sigma_0^+}$. The quantum equations however exhibit different behavior. Below
${\sigma_0^+}$ one still has (\call{ldvm}), and the solution above
${\sigma_0^+}$ can be computed in a Taylor expansion in (${\sigma^+}
- {\sigma_0^+}$). Defining ${\Sigma}({\sigma^-}) = {\partial_+}{\phi}
({\sigma_0^+}, {\sigma_-})$, equation (\call{pheq}) becomes
a simple equation for ${\Sigma}$:
$$
{8}{P}{\partial_-}{\Sigma} = {-} {P^{\prime}}({4}{\partial_-}{\phi}
{\Sigma} + {\lambda^2}{e^{2\rho}}),\eqno(sgeq)
$$
\noindent where (\call{ldvm}) should be substituted for the values
of ${\phi}, {\rho}$ along $({\sigma_0^+}, {\sigma^-})$. This is
easily integrated to yield
$$
{\Sigma}({\sigma^-}) = {{1}\over{2}} ( {{{M}\over{\sqrt{{P}({\sigma_0^+},
{\sigma^-})}}}} {-} {\lambda}),\eqno(sgsl)
$$
\noindent just above the shock wave.
The integration constant here is fixed by requiring that
asymptotically as ${\sigma^-}{\rightarrow}{-\infty}$ (on ${\cal I}_R^-)$
${\Sigma}$ agree with the classical ${f}$ shock wave solution.

For ${N}<{24}$, ${P}$ has no zeros, and ${\Sigma}$ is perfectly finite.
(For ${N}{\geq}{24}$, ${\Sigma}$ diverges at a finite value of ${\sigma^-}$.)
As explained in [\cite{rst}], an apparent horizon occurs whenever ${\Sigma}$
vanishes, and one may say that an ``apparent black hole'' has formed.
Since the minimum value of ${P}$ (at ${\phi} = {\phi_c})$ is
${{N}\over{24}} ({1} {-} {{N}\over{24}})$
there is no apparent horizon for sufficiently weak shock waves,
i.e., small ${M}$. For ${M} = {\lambda}
{\sqrt{{{N}\over{24}} ({1} - {{N}\over{24}})}}$, ${\Sigma}$
has a double zero where the shock wave crosses ${\phi} = {\phi_c}$, which
splits into two apparent horizons (containing a region of trapped
points\footnote{${\dagger}$}{A trapped point is one point for which
${\phi}$ increases along both outgoing null geodesics. This corresponds
to the four dimensional definition of a trapped surface when
${e^{-2\phi}}$ is interpreted as the size of the two spheres
[\cite{rst,ban,dxbh}].})
as ${M}$ increases. Since ${P}$ approaches ${{N}\over{24}}$ as ${\sigma^-}
{\rightarrow}{\infty}$ (on ${\cal I}_L^+)$, for ${M}>{\lambda}
{\sqrt{{{N}\over{24}}}}$
the region of trapped points extends all the way from the first apparent
horizon up to ${\cal I}_L^+$ along the ${f}$ shock wave.

It is of interest to determine the fate of the apparent black hole--or,
equivalently the region of trapped points--above the shock wave. This
is a difficult problem in general, but some progress on it can be made as
follows. At the apparent horizon, where ${\partial_+}{\phi} = {0}$,
equation (\call{pheq}) reduces to
$$
{\partial_-}({\partial_+}{\phi}) = {{-}} {{{P^{\prime}}{\lambda^2}
{e^{2\rho}}}\over{8P}}\eqno(abc)
$$
\noindent where ${P^{\prime}}$ is negative (positive) in the weak (strong)
coupling region where ${\phi}<{\phi_c} ~({\phi}>{\phi_c})$. It follows
that if one moves across an outer (inner) apparent horizon in the
direction of increasing ${\sigma^-}$, one always enters (leaves) the
interior of the apparent black hole where ${\partial_+}{\phi}>{0}$.

Thus if the outer apparent horizon (in the weak coupling region) is
followed above the shock wave, ${\sigma^+}$ will monotonically
increase until (or unless) it meets the line ${\phi} = {\phi_c}$
(which is spacelike inside the apparent horizon). If it does meet
this line, the boundary (now the inner horizon) must subsequently
continue along decreasing ${\sigma^+}$, back toward the shock
wave. Thus the apparent black hole ceases to exist for ${\sigma^+}$
greater than the value at which the apparent horizon meets
${\phi_c}$.\footnote{*}{Though it may (and will for ${M}>{\lambda}
{\sqrt{{N}\over{24}}})$ exist for large values of ${\sigma^-}$.}

It is thus crucial to determine whether or not the apparent horizon
actually reaches ${\phi_c}$. Following [\cite{rst}], the outer horizon may
be parameterized as the curve ${\hat{\sigma^-}}({\sigma^+})$.  Russo,
Susskind and Thorlacius [\cite{rst}] derive several formulae for
${\hat{\sigma^-}}$ from the condition
$${{d}\over{{d}{\sigma^+}}}\,\,{\partial_+}{\phi}({\hat\sigma}) =
{\partial_+}^2
{\phi}({\hat\sigma})+{{{d}{\hat{\sigma^-}}}\over{{d}{\sigma^+}}}
 {\partial_+}{\partial_-} {\phi}({\hat\sigma})={0}.\eqno(cdt)$$  For
the present system these become $$\eqalignno{
{{{d}{\hat{\sigma^-}}}\over{{d}{\sigma^+}}}&= {-}
{{{\partial^2_+}{\phi}}\over{{\partial_+}{\partial_-}{\phi}}}\cr &=
{{{4}{e^{2\phi-2\rho}}\,{P}}\over{{\lambda^2}{P^{\prime}}}}\,\,
{T^Q_{++}}&({aht})\cr} $$
\noindent where
$$
{T_{++}^{Q}} = {{N}\over{12}}({\partial_+}{\rho}{\partial_+}{\rho}{-}
{\partial_+^2}{\rho})
+{2}({\partial_+}({\rho}-{\phi})
{\partial_+}({\rho}-{\phi}) - {\partial_+^2}
({\rho}-{\phi})) + {t_+}\eqno(tqu)
$$
\noindent can be thought of as minus the energy in Hawking radiation
leaving the apparent black hole.
The second relation in (\call{aht}) implies that if ${T^Q_{++}}$ is always
negative (i.e. the black hole is evaporating), the outer apparent horizon is
timelike and will tend to
meet the spacelike line ${\phi} = {\phi_c}$.\footnote{${\dagger}$}
{If ${T^Q}$ goes to zero sufficiently fast the horizon could be asymptotically
null and avoid ${\phi_c}$.}
However we have unfortunately been unable to prove that ${T^Q_{++}}$ is
indeed everywhere negative.

Further progress can be made by considering a very small apparent black
hole, for which ${M}$ is just over the threshold for production of an
apparent horizon along the shock wave. In that case the apparent black
hole is formed in a small neighborhood of ${\phi_c}$, and its
evolution can be determined by a Taylor expansion about the point
$({\sigma^+_0}, {\sigma^-_c}){~~}{\equiv}{~~}
({\sigma^+_0}, {\sigma^+_0}\,\,
{-} {{1}\over{\lambda}}\,{\ell}{n}\,{{N}\over{24}})$
where the shock wave crosses the line ${\phi} = {\phi_c}$.
Using the first equation in (\call{aht}) and (\call{abc}) one finds that
the trajectory of the horizon is determined by
$$
{{{d}{\hat{\sigma^-}}}\over{{d}{\sigma^+}}}{~~}{\approx}{~~}
{{{2} ({{24}\over{N}} - {1})
{\partial^2_+}{\phi}({\sigma_0^+,}{\sigma_c^-})}\over
{{\lambda^3}({\sigma^-} - {\sigma_c^-})}}
{~~}{\equiv}{~~}{{-}}\,\,{{k}\over{{\sigma^-}-{\sigma^-_c}}}\eqno(sbh)
$$
One then finds for
${\sigma^+}>{\sigma^+_0}$
$$
{\hat{\sigma^-}} = {\sqrt{{2k}({\sigma^+_0}-{\sigma^+}) + {c^2}}} +
{\sigma^-_c}\eqno(trj)
$$
\noindent where $({\sigma^+_0}, {\sigma^-_c} {\pm}{c})$ are the initial
coordinates
of the inner and outer horizons along the shock wave. The behavior
of the trajectory depends crucially on the sign of
${\partial^2_+}{\phi}({\sigma_0^+,}{\sigma_c^-})$ or ${k}$.
According to (\call{trj}), if $k$ is positive the apparent black
hole shrinks and disappears at ${\sigma^+} = {\sigma^+_0} + {{c^2}\over{2k}}$,
soon after formation. On the other hand if $k$ is negative the apparent black
hole will initially grow in size,
and perturbation theory about the shock wave can not be used to determine its
ultimate fate.
The sign of $k$ is determined by continuing to one higher order the Taylor
expansion
about ${\sigma_0^+}$ used to find ${\Sigma}({\sigma^-})$. A tedious
computation reveals that, rather curiously, ${k}$ is positive for some values
of $N$ and negative for others.
We do not understand the significance of this. Perhaps the theory depends
qualitatively on $N$ even within the
range $0<N<24$.

It is of interest to study the behavior of the quantum stress tensor
${T^Q_{++}}$ of (\call{tqu}) along the shock wave.
At ${\sigma^+} = {\sigma_0^+}$,
${\partial_+}{\phi}$ (as well as ${\partial_+}{\rho}$) is discontinuous.
${T_{++}^Q}$ will therefore have a delta function at ${\sigma_0^+}$.
One easily finds
$$
{T_{++}^Q} ({\sigma_0^+}, {\sigma^-}) = {M}{\delta}({\sigma^+} - {\sigma_0^+})
({{e^{\lambda({\sigma_0^+} - {\sigma^-})}}\over{\sqrt{P}}} - {1})\eqno(eas)
$$
\noindent As the shock wave enters from ${\cal I}_R^- ({\sigma^-} =
{-\infty})$,
${T_{++}^Q}$ vanishes. As it moves in, ${P}$ increases, and negative
energy quantum fluctuations begin to accumulate along the shock wave. This
energy approaches a constant up on ${\cal I}^+_L$ which obeys
$$
{T_{++}^Q} {+} {T_{++}^f} = {0}\eqno(tqf)
$$
\noindent at $({\sigma_0^+}, {\infty})$. Thus the ${f}$ shock wave is
gravitationally cloaked by
a cloud of quantum fluctuations and arrives at ${\cal I}_L^+$ as
a ``zero-energy bound state.''\footnote{*}{This greatly strengthens
the analogy made in [\cite{sth,alf}] to the Schwinger model with a
position-dependent mass.}

This is in line with the conjecture made in [\cite{Evan}] that the
incoming state from ${\cal I}_R^-$ evaporates before forming a black hole, and
arrives as a zero energy bound state on ${\cal I}_L^+$. It is tempting to
speculate that this conjecture, though disproved for ${N}>{24}$ [\cite
{rst,ban}], might be applicable to ${N}<{24}$. However it is not clear
if this is consistent with the behavior of the apparent horizons.

We hope to answer these questions in the near future. For the moment
some further
insight can be gained by investigating static black hole solutions of
the type found for ${N}{>}{24}$ in [\cite{hwk,qbh}]. Following [\cite{qbh}]
this is best accomplished in terms of the variable
$$
{s} = {\lambda^{-2}}
{e^{{\lambda}{(\sigma^+}-{\sigma^-)}}}\eqno(surb)
$$
\noindent which vanishes on the horizon, is spacelike and positive outside
and timelike and negative inside. The equations of motion (\call{pheq})--
(\call{rheq}) for fields depending only on ${s}$ become
$$\eqalignno{
{8}{P} ({s}{\ddot\phi} + {\dot\phi})&= {P^{\prime}} ({\lambda^2}
{e^{2\rho}} - {4}{s}{\dot\phi}^2),&({speq})\cr
{2P}({s}{\ddot\rho} + {\dot\rho})&= {-} {e^{-4\phi}}
({\lambda^2}{e^{2\rho}} - {4s}{\dot\phi}^2),&({sreq})\cr}
$$
\noindent while the constraint is
$$
{e^{-2\phi}}(4{\dot\phi}{\dot\rho}-{2}{\ddot\phi})=
{{N}\over{12}}\,\,({\dot\rho}^2{-}{\ddot\rho})
{-} {2} (({\dot\rho} - {\dot\phi})^2 - {\ddot\rho} +
{\ddot\phi}) + {{\hat t}\over{s^2}},\eqno(sceq)
$$
\noindent where ``$^.$'' denotes differentiation with respect to ${s}$
and ${\hat t}$ is a constant. Finiteness of ${\ddot\phi}$ and ${\ddot\rho}$
at the horizon gives constraints for initial data at ${s} = {0}$:
$$\eqalignno{
{\dot\rho} ({0})&= {-}\,\,{{{\lambda^2}{e^{2\rho(0)-{4}{\phi}({0})}}}
\over
{{2}{P}({0})}},\cr
{\dot\phi}({0})&= {{{\lambda^2}{e^{2\rho(0)}}{P^{\prime}}(0)}\over
{8P(0)}},\cr
{\hat t}&= {0}.&({hred})\cr}
$$
\noindent Since ${\rho}(0)$ can be set to zero by a global coordinate
transformation, there is a one parameter family of inequivalent solutions
labelled by ${\phi}(0)$, or equivalently, the black hole mass.

The behavior of these solutions depends on whether ${\phi}(0)$ is less
than or greater than the critical value, ${\phi_c} = {{1}\over{2}}\,{\ell}{n}
\,\,{{24}\over{N}}$,
where
${P^{\prime}}$ changes sign. For large negative ${\phi}({0})$
(corresponding to large black holes), the solutions will differ little
outside the horizon from those found in [\cite{hwk,qbh}]. Asymptotically
the solution approaches the linear dilaton vacuum, but with a linearly
divergent ADM mass corresponding to the infinite radiation bath
required to stabilize the black hole against Hawking decay. As ${\phi}
(0)$ approaches ${\phi_c}$ the solutions begin to differ. This can be seen
from the fact that at ${\phi}({0}) = {\phi_c}$ the solution is exactly given by
(in the gauge ${\rho}({0}) = {0}$)
$$\eqalignno{
{\phi}&={{1}\over{2}}\,{\ell}{n}\,\,{{24}\over{N}}{~~}{\equiv}{~~}{\phi_c},\cr
{\rho}&={-}{\ell}{n}({1} + {{s}\over{\alpha}}).&({dstr})\cr}
$$
\noindent where ${\alpha} = {2}{\lambda^{-2}} ({{24}\over{N}} - {1})$.
This corresponds to deSitter space filled with Hawking
radiation. We presume that as ${\phi}(0)$ approaches ${\phi_c}$,
there is a growing deSitter like region outside the black hole. At
${\phi} (0) = {\phi_c}$, this region engulfs the entire spacetime and
the black hole horizon becomes a deSitter horizon.

One expects that a large slowly evaporating black hole is approximated
within some region by these static solutions with a slowly increasing
${\phi}(0)$. However one should not conclude from the above that
the endpoint of an apparent black hole formed by a massive incoming shock
wave is deSitter space. In that situation, the spacetime is always
asymptotic to the linear dilaton vacuum with a finite mass. One possibility
is that it looks like deSitter space within some region which then decays
back to the linear dilaton vacuum.

Inside the horizon, the ghost modifications have a crucial effect even for
${\phi}(0)<<{\phi_c}$. When ${N}>{24}$, it was shown that ${\phi}$
increases (now in a timelike direction) until a zero of ${P}$ and a singularity
is reached. The resulting spacetime has a causal structure identical
to that of the static classical black hole solution. However for ${N}<
{24}$ there are no zeros of ${P}$. In this case one finds from (\call{speq})
that for ${\phi}<{\phi_c}$ both ${\phi}$ and ${s}{\dot\phi}$
are initially increasing as before. ${\phi}$ will then inevitably cross
${\phi_c}$,
at which point ${s}{\dot\phi}$ must start to decrease, and ${\phi}$
``bounces'' off of some maximum value rather than becoming singular.
This behavior can be understood from a linearized analysis for small
${\phi}(0)$. To leading order the ${\rho}$ equation is
independent of ${\phi}$ and yields the deSitter solution (\call{dstr}). The
linearization of (\call{pheq}) is then
$$
{s}{\ddot\phi} + {\dot\phi} = {{2\alpha}\over{({s}+{\alpha})^2}}\,{\phi}.
\eqno(phln)
$$
\noindent Define the new timelike variable
$$
{u} = {\ell}{n} [{{{\sqrt{\alpha}} + {\sqrt{-s}}}\over
{{\sqrt{\alpha}} - {\sqrt{-s}}}}]\eqno(udet)
$$
\noindent which runs from zero at the horizon  to plus infinity at
future timelike infinity  $({s} = {-}{\alpha})$. The linearized ${\phi}$
equation then becomes
$$
{\partial_u^2}{\phi} + {\cot}{\rm h}{\it u}
{\partial_u}{\phi} + {2}{\phi} = {0},
\eqno(uphg)
$$
\noindent while the boundary condition (\call{hred}) implies
${\partial_u}{\phi}$
vanishes at the horizon.
As the coefficient of the first derivative term is positive for
${u}>{0}$, this is a damped harmonic oscillator. Thus excursions of
${\phi}$ are damped inside the horizon and the linearized approximation
does not break down. We therefore conclude that, at least for small
${\phi}(0)$, the interior of the black hole is non-singular
and asymptotic to deSitter
space, as illustrated in Figure 2.

Evidently this system is very resistant
to singularity formation: even if small black hole is forced into existence
by continuously pumping in energy from infinity, there is no singularity
in its interior.

For large ${\phi}(0)$ the equations are harder to analyze, but we expect
similar behavior. Numerical work on this question is in progress
[\cite{bist}].

Clearly this set of equations exhibits complex and unusual behavior
that we do not yet fully understand, and which merits further
investigation. Our preliminary investigations have failed to uncover
any black hole type singularities, but their existence is certainly
not ruled out. We also do not know if the equations give a qualitatively
correct description of ${N}<{24}$ matter fields coupled to dilaton
gravity, because higher loop corrections could be important.
What {\it has} been established, however, is that the
nature of the black hole formation/evaporation process, including the
singularity structure, depends qualitatively on the properties of the
matter sector. It is an urgent problem to characterize the possible
behaviors.

It is intriguing that the outcome of two-dimensional gravitational collapse
depends qualitatively on the matter content of the universe. Perhaps
this will also turn out to be true in four dimensions, and lead
to constraints on the spectrum of elementary particles.
\bigskip
\centerline{\bf Acknowledgements}

I am grateful to B. Birnir, S. Giddings, J. Harvey, L. Susskind and
L. Thorlacius for useful discussions. This work was supported in part
by DOE-91ER40618.

\references

\refis{Hawk} S. W. Hawking, {\it Particle Creation by Black Holes},
Comm. Math. Phys. {\bf 43} (1975), 199.

\refis{Evan} C. G. Callan, S. B. Giddings, J. A. Harvey and A. Strominger,
{\it Evanescent Black Holes}, Phys. Rev. D {\bf 45}, R1005 (1992).

\refis{ban} T. Banks, A. Dabholkar, M. R. Douglas and M. O'Loughlin, {\it
Are Horned Particles the Climax of Hawking Evaporation?} Rutgers University
preprint, RU-91-54, January 1992.

\refis{rst} J. G. Russo, L. Susskind and L. Thorlacius, {\it Black
Hole Evaporation in ${1} + {1}$ Dimensions}, Stanford University preprint,
SU-ITP-92-4, January 1992.

\refis{qbh} B. Birnir, S. B. Giddings, J. A. Harvey and A. Strominger, {\it
Quantum Black Holes}, preprint UCSB-TH-92-08, EFI-92-16, hepth@xxx/9203032.

\refis{hwk} S. W. Hawking, {\it Evaporation of Two Dimensional Black Holes},
Caltech preprint, CALT-68-\#1, hepth@xxx/9203054.

\refis{sth} L. Susskind and L. Thorlacius, {\it Hawking Radiation
and Backreaction}, Stanford preprint, SU-ITP-92-12, hepth@xxx/9203052.

\refis{Witt} E. Witten, {\it Phys. Rev.} {\bf D44} (1991), 314.

\refis{alf} M. Alford and A. Strominger, {\it S-Wave Scattering of
Charged Fermions by a Magnetic Black Hole}, preprint NSF-ITP-92-13, February
1992.

\refis{hver} J. G. Russo and A. A. Tseytlin, {\it Scalar-Tensor Quantum Gravity
in Two Dimensions} preprint SU-ITP-92-2, DAMTP-1-1992; H. Verlinde
{\it Black
Holes and Strings in Two Dimensions} Princeton preprint PUPT-1303,
to appear in the proceedings of the Sixth Marcel Grossman Meeting.

\refis{bist} B. Birnir, A. Strominger, L. Thorlacius, ... in progress.

\refis{dxbh} S. B. Giddings and A. Strominger, {\it Dynamics of Extremal
Black Holes}, Phys. Rev. D, to appear (1992).

\endreferences

\endit

%%Figure 1 begins here.
%!PS-Adobe-2.0 EPSF-1.2
%%Creator: idraw
%%DocumentFonts: Times-Roman Times-Italic Helvetica Times-Bold
%%Pages: 1
%%BoundingBox: 80 153 534 621
%%EndComments

/arrowHeight 11 def
/arrowWidth 6 def

/IdrawDict 54 dict def
IdrawDict begin

/reencodeISO {
dup dup findfont dup length dict begin
{ 1 index /FID ne { def }{ pop pop } ifelse } forall
/Encoding ISOLatin1Encoding def
currentdict end definefont
} def

/ISOLatin1Encoding [
/.notdef/.notdef/.notdef/.notdef/.notdef/.notdef/.notdef/.notdef
/.notdef/.notdef/.notdef/.notdef/.notdef/.notdef/.notdef/.notdef
/.notdef/.notdef/.notdef/.notdef/.notdef/.notdef/.notdef/.notdef
/.notdef/.notdef/.notdef/.notdef/.notdef/.notdef/.notdef/.notdef
/space/exclam/quotedbl/numbersign/dollar/percent/ampersand/quoteright
/parenleft/parenright/asterisk/plus/comma/minus/period/slash
/zero/one/two/three/four/five/six/seven/eight/nine/colon/semicolon
/less/equal/greater/question/at/A/B/C/D/E/F/G/H/I/J/K/L/M/N
/O/P/Q/R/S/T/U/V/W/X/Y/Z/bracketleft/backslash/bracketright
/asciicircum/underscore/quoteleft/a/b/c/d/e/f/g/h/i/j/k/l/m
/n/o/p/q/r/s/t/u/v/w/x/y/z/braceleft/bar/braceright/asciitilde
/.notdef/.notdef/.notdef/.notdef/.notdef/.notdef/.notdef/.notdef
/.notdef/.notdef/.notdef/.notdef/.notdef/.notdef/.notdef/.notdef
/.notdef/dotlessi/grave/acute/circumflex/tilde/macron/breve
/dotaccent/dieresis/.notdef/ring/cedilla/.notdef/hungarumlaut
/ogonek/caron/space/exclamdown/cent/sterling/currency/yen/brokenbar
/section/dieresis/copyright/ordfeminine/guillemotleft/logicalnot
/hyphen/registered/macron/degree/plusminus/twosuperior/threesuperior
/acute/mu/paragraph/periodcentered/cedilla/onesuperior/ordmasculine
/guillemotright/onequarter/onehalf/threequarters/questiondown
/Agrave/Aacute/Acircumflex/Atilde/Adieresis/Aring/AE/Ccedilla
/Egrave/Eacute/Ecircumflex/Edieresis/Igrave/Iacute/Icircumflex
/Idieresis/Eth/Ntilde/Ograve/Oacute/Ocircumflex/Otilde/Odieresis
/multiply/Oslash/Ugrave/Uacute/Ucircumflex/Udieresis/Yacute
/Thorn/germandbls/agrave/aacute/acircumflex/atilde/adieresis
/aring/ae/ccedilla/egrave/eacute/ecircumflex/edieresis/igrave
/iacute/icircumflex/idieresis/eth/ntilde/ograve/oacute/ocircumflex
/otilde/odieresis/divide/oslash/ugrave/uacute/ucircumflex/udieresis
/yacute/thorn/ydieresis
] def
/Times-Roman reencodeISO def
/Times-Italic reencodeISO def
/Helvetica reencodeISO def
/Times-Bold reencodeISO def

/none null def
/numGraphicParameters 17 def
/stringLimit 65535 def

/Begin {
save
numGraphicParameters dict begin
} def

/End {
end
restore
} def

/SetB {
dup type /nulltype eq {
pop
false /brushRightArrow idef
false /brushLeftArrow idef
true /brushNone idef
} {
/brushDashOffset idef
/brushDashArray idef
0 ne /brushRightArrow idef
0 ne /brushLeftArrow idef
/brushWidth idef
false /brushNone idef
} ifelse
} def

/SetCFg {
/fgblue idef
/fggreen idef
/fgred idef
} def

/SetCBg {
/bgblue idef
/bggreen idef
/bgred idef
} def

/SetF {
/printSize idef
/printFont idef
} def

/SetP {
dup type /nulltype eq {
pop true /patternNone idef
} {
dup -1 eq {
/patternGrayLevel idef
/patternString idef
} {
/patternGrayLevel idef
} ifelse
false /patternNone idef
} ifelse
} def

/BSpl {
0 begin
storexyn
newpath
n 1 gt {
0 0 0 0 0 0 1 1 true subspline
n 2 gt {
0 0 0 0 1 1 2 2 false subspline
1 1 n 3 sub {
/i exch def
i 1 sub dup i dup i 1 add dup i 2 add dup false subspline
} for
n 3 sub dup n 2 sub dup n 1 sub dup 2 copy false subspline
} if
n 2 sub dup n 1 sub dup 2 copy 2 copy false subspline
patternNone not brushLeftArrow not brushRightArrow not and and { ifill } if
brushNone not { istroke } if
0 0 1 1 leftarrow
n 2 sub dup n 1 sub dup rightarrow
} if
end
} dup 0 4 dict put def

/Circ {
newpath
0 360 arc
patternNone not { ifill } if
brushNone not { istroke } if
} def

/CBSpl {
0 begin
dup 2 gt {
storexyn
newpath
n 1 sub dup 0 0 1 1 2 2 true subspline
1 1 n 3 sub {
/i exch def
i 1 sub dup i dup i 1 add dup i 2 add dup false subspline
} for
n 3 sub dup n 2 sub dup n 1 sub dup 0 0 false subspline
n 2 sub dup n 1 sub dup 0 0 1 1 false subspline
patternNone not { ifill } if
brushNone not { istroke } if
} {
Poly
} ifelse
end
} dup 0 4 dict put def

/Elli {
0 begin
newpath
4 2 roll
translate
scale
0 0 1 0 360 arc
patternNone not { ifill } if
brushNone not { istroke } if
end
} dup 0 1 dict put def

/Line {
0 begin
2 storexyn
newpath
x 0 get y 0 get moveto
x 1 get y 1 get lineto
brushNone not { istroke } if
0 0 1 1 leftarrow
0 0 1 1 rightarrow
end
} dup 0 4 dict put def

/MLine {
0 begin
storexyn
newpath
n 1 gt {
x 0 get y 0 get moveto
1 1 n 1 sub {
/i exch def
x i get y i get lineto
} for
patternNone not brushLeftArrow not brushRightArrow not and and { ifill } if
brushNone not { istroke } if
0 0 1 1 leftarrow
n 2 sub dup n 1 sub dup rightarrow
} if
end
} dup 0 4 dict put def

/Poly {
3 1 roll
newpath
moveto
-1 add
{ lineto } repeat
closepath
patternNone not { ifill } if
brushNone not { istroke } if
} def

/Rect {
0 begin
/t exch def
/r exch def
/b exch def
/l exch def
newpath
l b moveto
l t lineto
r t lineto
r b lineto
closepath
patternNone not { ifill } if
brushNone not { istroke } if
end
} dup 0 4 dict put def

/Text {
ishow
} def

/idef {
dup where { pop pop pop } { exch def } ifelse
} def

/ifill {
0 begin
gsave
patternGrayLevel -1 ne {
fgred bgred fgred sub patternGrayLevel mul add
fggreen bggreen fggreen sub patternGrayLevel mul add
fgblue bgblue fgblue sub patternGrayLevel mul add setrgbcolor
eofill
} {
eoclip
originalCTM setmatrix
pathbbox /t exch def /r exch def /b exch def /l exch def
/w r l sub ceiling cvi def
/h t b sub ceiling cvi def
/imageByteWidth w 8 div ceiling cvi def
/imageHeight h def
bgred bggreen bgblue setrgbcolor
eofill
fgred fggreen fgblue setrgbcolor
w 0 gt h 0 gt and {
l b translate w h scale
w h true [w 0 0 h neg 0 h] { patternproc } imagemask
} if
} ifelse
grestore
end
} dup 0 8 dict put def

/istroke {
gsave
brushDashOffset -1 eq {
[] 0 setdash
1 setgray
} {
brushDashArray brushDashOffset setdash
fgred fggreen fgblue setrgbcolor
} ifelse
brushWidth setlinewidth
originalCTM setmatrix
stroke
grestore
} def

/ishow {
0 begin
gsave
fgred fggreen fgblue setrgbcolor
/fontDict printFont printSize scalefont dup setfont def
/descender fontDict begin 0 [FontBBox] 1 get FontMatrix end
transform exch pop def
/vertoffset 1 printSize sub descender sub def {
0 vertoffset moveto show
/vertoffset vertoffset printSize sub def
} forall
grestore
end
} dup 0 3 dict put def
/patternproc {
0 begin
/patternByteLength patternString length def
/patternHeight patternByteLength 8 mul sqrt cvi def
/patternWidth patternHeight def
/patternByteWidth patternWidth 8 idiv def
/imageByteMaxLength imageByteWidth imageHeight mul
stringLimit patternByteWidth sub min def
/imageMaxHeight imageByteMaxLength imageByteWidth idiv patternHeight idiv
patternHeight mul patternHeight max def
/imageHeight imageHeight imageMaxHeight sub store
/imageString imageByteWidth imageMaxHeight mul patternByteWidth add string def
0 1 imageMaxHeight 1 sub {
/y exch def
/patternRow y patternByteWidth mul patternByteLength mod def
/patternRowString patternString patternRow patternByteWidth getinterval def
/imageRow y imageByteWidth mul def
0 patternByteWidth imageByteWidth 1 sub {
/x exch def
imageString imageRow x add patternRowString putinterval
} for
} for
imageString
end
} dup 0 12 dict put def

/min {
dup 3 2 roll dup 4 3 roll lt { exch } if pop
} def

/max {
dup 3 2 roll dup 4 3 roll gt { exch } if pop
} def

/midpoint {
0 begin
/y1 exch def
/x1 exch def
/y0 exch def
/x0 exch def
x0 x1 add 2 div
y0 y1 add 2 div
end
} dup 0 4 dict put def

/thirdpoint {
0 begin
/y1 exch def
/x1 exch def
/y0 exch def
/x0 exch def
x0 2 mul x1 add 3 div
y0 2 mul y1 add 3 div
end
} dup 0 4 dict put def

/subspline {
0 begin
/movetoNeeded exch def
y exch get /y3 exch def
x exch get /x3 exch def
y exch get /y2 exch def
x exch get /x2 exch def
y exch get /y1 exch def
x exch get /x1 exch def
y exch get /y0 exch def
x exch get /x0 exch def
x1 y1 x2 y2 thirdpoint
/p1y exch def
/p1x exch def
x2 y2 x1 y1 thirdpoint
/p2y exch def
/p2x exch def
x1 y1 x0 y0 thirdpoint
p1x p1y midpoint
/p0y exch def
/p0x exch def
x2 y2 x3 y3 thirdpoint
p2x p2y midpoint
/p3y exch def
/p3x exch def
movetoNeeded { p0x p0y moveto } if
p1x p1y p2x p2y p3x p3y curveto
end
} dup 0 17 dict put def

/storexyn {
/n exch def
/y n array def
/x n array def
n 1 sub -1 0 {
/i exch def
y i 3 2 roll put
x i 3 2 roll put
} for
} def

/SSten {
fgred fggreen fgblue setrgbcolor
dup true exch 1 0 0 -1 0 6 -1 roll matrix astore
} def

/FSten {
dup 3 -1 roll dup 4 1 roll exch
newpath
0 0 moveto
dup 0 exch lineto
exch dup 3 1 roll exch lineto
0 lineto
closepath
bgred bggreen bgblue setrgbcolor
eofill
SSten
} def

/Rast {
exch dup 3 1 roll 1 0 0 -1 0 6 -1 roll matrix astore
} def

%%EndProlog

%%BeginIdrawPrologue
/arrowhead {
0 begin
transform originalCTM itransform
/taily exch def
/tailx exch def
transform originalCTM itransform
/tipy exch def
/tipx exch def
/dy tipy taily sub def
/dx tipx tailx sub def
/angle dx 0 ne dy 0 ne or { dy dx atan } { 90 } ifelse def
gsave
originalCTM setmatrix
tipx tipy translate
angle rotate
newpath
arrowHeight neg arrowWidth 2 div moveto
0 0 lineto
arrowHeight neg arrowWidth 2 div neg lineto
patternNone not {
originalCTM setmatrix
/padtip arrowHeight 2 exp 0.25 arrowWidth 2 exp mul add sqrt brushWidth mul
arrowWidth div def
/padtail brushWidth 2 div def
tipx tipy translate
angle rotate
padtip 0 translate
arrowHeight padtip add padtail add arrowHeight div dup scale
arrowheadpath
ifill
} if
brushNone not {
originalCTM setmatrix
tipx tipy translate
angle rotate
arrowheadpath
istroke
} if
grestore
end
} dup 0 9 dict put def

/arrowheadpath {
newpath
arrowHeight neg arrowWidth 2 div moveto
0 0 lineto
arrowHeight neg arrowWidth 2 div neg lineto
} def

/leftarrow {
0 begin
y exch get /taily exch def
x exch get /tailx exch def
y exch get /tipy exch def
x exch get /tipx exch def
brushLeftArrow { tipx tipy tailx taily arrowhead } if
end
} dup 0 4 dict put def

/rightarrow {
0 begin
y exch get /tipy exch def
x exch get /tipx exch def
y exch get /taily exch def
x exch get /tailx exch def
brushRightArrow { tipx tipy tailx taily arrowhead } if
end
} dup 0 4 dict put def

%%EndIdrawPrologue

%I Idraw 10 Grid 10 10

%%Page: 1 1

Begin
%I b u
%I cfg u
%I cbg u
%I f u
%I p u
%I t
[ 0.866142 0 0 0.866142 0 0 ] concat
/originalCTM matrix currentmatrix def

Begin %I Text
%I cfg Black
0 0 0 SetCFg
%I f *-times-medium-r-*-120-*
Times-Roman 12 SetF
%I t
[ 1 0 0 1 451.352 497.785 ] concat
%I
[
(+)
] Text
End

Begin %I Text
%I cfg Black
0 0 0 SetCFg
%I f *-times-medium-r-*-140-*
Times-Roman 14 SetF
%I t
[ 0.696146 -0.7179 0.7179 0.696146 361.703 569.604 ] concat
%I
[
(f-wave)
] Text
End

Begin %I Text
%I cfg Black
0 0 0 SetCFg
%I f *-times-medium-r-*-120-*
Times-Roman 12 SetF
%I t
[ 1 0 0 1 170.352 684.785 ] concat
%I
[
(+)
] Text
End

Begin %I Text
%I cfg Black
0 0 0 SetCFg
%I f *-times-medium-r-*-120-*
Times-Roman 12 SetF
%I t
[ 1 0 0 1 203.352 463.285 ] concat
%I
[
(-)
] Text
End

Begin %I Text
%I cfg Black
0 0 0 SetCFg
%I f *-times-medium-r-*-120-*
Times-Roman 12 SetF
%I t
[ 1 0 0 1 464.352 457.785 ] concat
%I
[
(-)
] Text
End

Begin %I Text
%I cfg Black
0 0 0 SetCFg
%I f *-times-medium-r-*-120-*
Times-Roman 12 SetF
%I t
[ 1 0 0 1 530.576 598.801 ] concat
%I
[
(R)
] Text
End

Begin %I Text
%I cfg Black
0 0 0 SetCFg
%I f *-times-medium-r-*-120-*
Times-Roman 12 SetF
%I t
[ 1 0 0 1 170.352 669.785 ] concat
%I
[
(L)
] Text
End

Begin %I Line
%I b 65535
1 0 1 [] 0 SetB
%I cfg Black
0 0 0 SetCFg
%I cbg White
1 1 1 SetCBg
%I p
0.5 SetP
%I t
[ 1 -0 -0 1 164.352 350.785 ] concat
%I
280 43 356 120 Line
%I 1
End

Begin %I Text
%I cfg Black
0 0 0 SetCFg
%I f *-times-medium-r-*-120-*
Times-Roman 12 SetF
%I t
[ 1 0 0 1 487.352 424.785 ] concat
%I
[
(+)
] Text
End

Begin %I Text
%I cfg Black
0 0 0 SetCFg
%I f *-times-medium-r-*-140-*
Times-Roman 14 SetF
%I t
[ 1 0 0 1 285.257 377.116 ] concat
%I
[
(-)
] Text
End

Begin %I Text
%I cfg Black
0 0 0 SetCFg
%I f *-times-medium-r-*-120-*
Times-Roman 12 SetF
%I t
[ 1 0 0 1 156.352 428.785 ] concat
%I
[
(-)
] Text
End

Begin %I Text
%I cfg Black
0 0 0 SetCFg
%I f *-times-medium-r-*-120-*
Times-Roman 12 SetF
%I t
[ 1 0 0 1 451.352 482.785 ] concat
%I
[
(0)
] Text
End

Begin %I Rect
%I b 65535
0 0 0 [] 0 SetB
%I cfg Black
0 0 0 SetCFg
%I cbg White
1 1 1 SetCBg
none SetP %I p n
%I t
[ 1 -0 -0 1 89.3523 303.785 ] concat
%I
173 410 173 411 Rect
End

Begin %I Text
%I cfg Black
0 0 0 SetCFg
%I f *-times-medium-r-*-120-*
Times-Roman 12 SetF
%I t
[ 1 0 0 1 201.352 447.785 ] concat
%I
[
(L)
] Text
End

Begin %I Text
%I cfg Black
0 0 0 SetCFg
%I f *-times-medium-i-*-140-*
Times-Italic 14 SetF
%I t
[ 1 0 0 1 196.352 458.785 ] concat
%I
[
(I)
] Text
End

Begin %I Text
%I cfg Black
0 0 0 SetCFg
%I f *-times-medium-i-*-140-*
Times-Italic 14 SetF
%I t
[ 1 0 0 1 164.352 678.785 ] concat
%I
[
(I)
] Text
End

Begin %I Text
%I cfg Black
0 0 0 SetCFg
%I f *-times-medium-r-*-120-*
Times-Roman 12 SetF
%I t
[ 1 0 0 1 534.374 615.482 ] concat
%I
[
(+)
] Text
End

Begin %I Text
%I cfg Black
0 0 0 SetCFg
%I f *-times-medium-i-*-140-*
Times-Italic 14 SetF
%I t
[ 1 0 0 1 525.238 606.97 ] concat
%I
[
(I)
] Text
End

Begin %I Text
%I cfg Black
0 0 0 SetCFg
%I f *-times-medium-r-*-120-*
Times-Roman 12 SetF
%I t
[ 1 0 0 1 463.352 443.785 ] concat
%I
[
(R)
] Text
End

Begin %I Text
%I cfg Black
0 0 0 SetCFg
%I f *-times-medium-i-*-140-*
Times-Italic 14 SetF
%I t
[ 1 0 0 1 458.352 452.785 ] concat
%I
[
(I)
] Text
End

Begin %I Line
%I b 65535
3 1 0 [] 0 SetB
%I cfg Black
0 0 0 SetCFg
%I cbg White
1 1 1 SetCBg
%I p
0.75 SetP
%I t
[ 0.25 -0 -0 0.25 365.477 430.597 ] concat
%I
118 461 318 261 Line
%I 4
End

Begin %I Line
%I b 65535
3 0 1 [] 0 SetB
%I cfg Black
0 0 0 SetCFg
%I cbg White
1 1 1 SetCBg
%I p
0.75 SetP
%I t
[ 0.25 -0 -0 0.25 291.477 519.597 ] concat
%I
414 103 247 273 Line
%I 4
End

Begin %I Line
%I b 65535
3 0 1 [] 0 SetB
%I cfg Black
0 0 0 SetCFg
%I cbg White
1 1 1 SetCBg
%I p
0.75 SetP
%I t
[ 0.25 -0 -0 0.25 292.308 519.597 ] concat
%I
248 270 71 449 Line
%I 4
End

Begin %I Line
%I b 65535
3 0 1 [] 0 SetB
%I cfg Black
0 0 0 SetCFg
%I cbg White
1 1 1 SetCBg
%I p
0.75 SetP
%I t
[ 0.25 -0 -0 0.25 217.227 599.847 ] concat
%I
367 126 216 280 Line
%I 4
End

Begin %I Line
%I b 65535
3 0 1 [] 0 SetB
%I cfg Black
0 0 0 SetCFg
%I cbg White
1 1 1 SetCBg
%I p
0.75 SetP
%I t
[ 0.125 -0 -0 0.125 212.102 660.347 ] concat
%I
473 76 236 313 Line
%I 8
End

Begin %I Text
%I cfg Black
0 0 0 SetCFg
%I f *-times-medium-r-*-140-*
Times-Roman 14 SetF
%I t
[ 1 0 0 1 234.186 539.167 ] concat
%I
[
(Dilaton)
] Text
End

Begin %I Line
%I b 65535
1 0 1 [] 0 SetB
%I cfg Black
0 0 0 SetCFg
%I cbg White
1 1 1 SetCBg
%I p
0.5 SetP
%I t
[ 0.25 -0 -0 0.25 106.227 343.785 ] concat
%I
382 198 110 471 Line
%I 4
End

Begin %I Text
%I cfg Black
0 0 0 SetCFg
%I f *-helvetica-medium-r-*-140-*
Helvetica 14 SetF
%I t
[ 1 0 0 1 184.352 593.785 ] concat
%I
[
(Linear)
] Text
End

Begin %I Text
%I cfg Black
0 0 0 SetCFg
%I f *-helvetica-medium-r-*-140-*
Helvetica 14 SetF
%I t
[ 1 0 0 1 288.352 496.785 ] concat
%I
[
(Vacuum)
] Text
End

Begin %I BSpl
%I b 65520
3 0 0 [12 4] 0 SetB
%I cfg Black
0 0 0 SetCFg
%I cbg White
1 1 1 SetCBg
none SetP %I p n
%I t
[ 0.830966 -0 -0 0.830966 -142.926 201.925 ] concat
%I 6
466 612
510 571
657 430
741 419
783 445
799 461
6 BSpl
%I 1
End

Begin %I MLine
%I b 65535
2 0 0 [] 0 SetB
%I cfg Black
0 0 0 SetCFg
%I cbg White
1 1 1 SetCBg
none SetP %I p n
%I t
[ 0.830966 -0 -0 0.830966 -142.926 201.925 ] concat
%I 5
465 612
307 453
558 202
807 452
798 461
5 MLine
%I 1
End

Begin %I Text
%I cfg Black
0 0 0 SetCFg
%I f *-helvetica-medium-r-*-140-*
Helvetica 14 SetF
%I t
[ 3.07692 0 0 3.07692 370.006 661.016 ] concat
%I
[
(?)
] Text
End

Begin %I Text
%I cfg Black
0 0 0 SetCFg
%I f *-helvetica-medium-r-*-120-*
Helvetica 12 SetF
%I t
[ 1 0 0 1 334.493 375.109 ] concat
%I
[
(-)
] Text
End

Begin %I Text
%I cfg Black
0 0 0 SetCFg
%I f *-helvetica-medium-r-*-120-*
Helvetica 12 SetF
%I t
[ 1 0 0 1 121.859 262.575 ] concat
%I
[
(Figure 1.  An f shock wave incident  on the linear dilaton vacuum.)
()
(For N<24, no singularities are encountered in a Taylor expansion)
()
(above the shock wave, but this expansion does not probe the region)
()
(above the dashed line. )
] Text
End

Begin %I Text
%I cfg Black
0 0 0 SetCFg
%I f *-times-bold-r-*-140-*
Times-Bold 14 SetF
%I t
[ 0.848528 -0.848528 0.848528 0.848528 149.345 425.906 ] concat
%I
[
(6)
] Text
End

Begin %I Text
%I cfg Black
0 0 0 SetCFg
%I f *-times-medium-r-*-140-*
Times-Roman 14 SetF
%I t
[ 0.848528 -0.848528 0.848528 0.848528 482.164 422.184 ] concat
%I
[
(6)
] Text
End

Begin %I Text
%I cfg Black
0 0 0 SetCFg
%I f *-times-medium-r-*-140-*
Times-Roman 14 SetF
%I t
[ 0.848528 -0.848528 0.848528 0.848528 446.123 493.337 ] concat
%I
[
(6)
] Text
End

Begin %I Text
%I cfg Black
0 0 0 SetCFg
%I f *-times-bold-r-*-140-*
Times-Bold 14 SetF
%I t
[ 1.2168e-08 -1 1 1.2168e-08 108.613 573.903 ] concat
%I
[
(8)
] Text
End

Begin %I Text
%I cfg Black
0 0 0 SetCFg
%I f *-times-bold-r-*-140-*
Times-Bold 14 SetF
%I t
[ 1.2168e-08 -1 1 1.2168e-08 550.378 574.425 ] concat
%I
[
(8)
] Text
End

Begin %I Text
%I cfg Black
0 0 0 SetCFg
%I f *-times-bold-r-*-140-*
Times-Bold 14 SetF
%I t
[ 1.2168e-08 -1 1 1.2168e-08 355.622 373.852 ] concat
%I
[
(8)
] Text
End

Begin %I Text
%I cfg Black
0 0 0 SetCFg
%I f *-times-bold-r-*-140-*
Times-Bold 14 SetF
%I t
[ 1.2168e-08 -1 1 1.2168e-08 307.947 374.374 ] concat
%I
[
(8)
] Text
End

End %I eop

showpage

%%Trailer

end

%Figure 2 begins here.

%!PS-Adobe-2.0 EPSF-1.2
%%Creator: idraw
%%DocumentFonts: Helvetica
%%Pages: 1
%%BoundingBox: 94 208 507 534
%%EndComments

/arrowHeight 11 def
/arrowWidth 6 def

/IdrawDict 51 dict def
IdrawDict begin

/reencodeISO {
dup dup findfont dup length dict begin
{ 1 index /FID ne { def }{ pop pop } ifelse } forall
/Encoding ISOLatin1Encoding def
currentdict end definefont
} def

/ISOLatin1Encoding [
/.notdef/.notdef/.notdef/.notdef/.notdef/.notdef/.notdef/.notdef
/.notdef/.notdef/.notdef/.notdef/.notdef/.notdef/.notdef/.notdef
/.notdef/.notdef/.notdef/.notdef/.notdef/.notdef/.notdef/.notdef
/.notdef/.notdef/.notdef/.notdef/.notdef/.notdef/.notdef/.notdef
/space/exclam/quotedbl/numbersign/dollar/percent/ampersand/quoteright
/parenleft/parenright/asterisk/plus/comma/minus/period/slash
/zero/one/two/three/four/five/six/seven/eight/nine/colon/semicolon
/less/equal/greater/question/at/A/B/C/D/E/F/G/H/I/J/K/L/M/N
/O/P/Q/R/S/T/U/V/W/X/Y/Z/bracketleft/backslash/bracketright
/asciicircum/underscore/quoteleft/a/b/c/d/e/f/g/h/i/j/k/l/m
/n/o/p/q/r/s/t/u/v/w/x/y/z/braceleft/bar/braceright/asciitilde
/.notdef/.notdef/.notdef/.notdef/.notdef/.notdef/.notdef/.notdef
/.notdef/.notdef/.notdef/.notdef/.notdef/.notdef/.notdef/.notdef
/.notdef/dotlessi/grave/acute/circumflex/tilde/macron/breve
/dotaccent/dieresis/.notdef/ring/cedilla/.notdef/hungarumlaut
/ogonek/caron/space/exclamdown/cent/sterling/currency/yen/brokenbar
/section/dieresis/copyright/ordfeminine/guillemotleft/logicalnot
/hyphen/registered/macron/degree/plusminus/twosuperior/threesuperior
/acute/mu/paragraph/periodcentered/cedilla/onesuperior/ordmasculine
/guillemotright/onequarter/onehalf/threequarters/questiondown
/Agrave/Aacute/Acircumflex/Atilde/Adieresis/Aring/AE/Ccedilla
/Egrave/Eacute/Ecircumflex/Edieresis/Igrave/Iacute/Icircumflex
/Idieresis/Eth/Ntilde/Ograve/Oacute/Ocircumflex/Otilde/Odieresis
/multiply/Oslash/Ugrave/Uacute/Ucircumflex/Udieresis/Yacute
/Thorn/germandbls/agrave/aacute/acircumflex/atilde/adieresis
/aring/ae/ccedilla/egrave/eacute/ecircumflex/edieresis/igrave
/iacute/icircumflex/idieresis/eth/ntilde/ograve/oacute/ocircumflex
/otilde/odieresis/divide/oslash/ugrave/uacute/ucircumflex/udieresis
/yacute/thorn/ydieresis
] def
/Helvetica reencodeISO def

/none null def
/numGraphicParameters 17 def
/stringLimit 65535 def

/Begin {
save
numGraphicParameters dict begin
} def

/End {
end
restore
} def

/SetB {
dup type /nulltype eq {
pop
false /brushRightArrow idef
false /brushLeftArrow idef
true /brushNone idef
} {
/brushDashOffset idef
/brushDashArray idef
0 ne /brushRightArrow idef
0 ne /brushLeftArrow idef
/brushWidth idef
false /brushNone idef
} ifelse
} def

/SetCFg {
/fgblue idef
/fggreen idef
/fgred idef
} def

/SetCBg {
/bgblue idef
/bggreen idef
/bgred idef
} def

/SetF {
/printSize idef
/printFont idef
} def

/SetP {
dup type /nulltype eq {
pop true /patternNone idef
} {
dup -1 eq {
/patternGrayLevel idef
/patternString idef
} {
/patternGrayLevel idef
} ifelse
false /patternNone idef
} ifelse
} def

/BSpl {
0 begin
storexyn
newpath
n 1 gt {
0 0 0 0 0 0 1 1 true subspline
n 2 gt {
0 0 0 0 1 1 2 2 false subspline
1 1 n 3 sub {
/i exch def
i 1 sub dup i dup i 1 add dup i 2 add dup false subspline
} for
n 3 sub dup n 2 sub dup n 1 sub dup 2 copy false subspline
} if
n 2 sub dup n 1 sub dup 2 copy 2 copy false subspline
patternNone not brushLeftArrow not brushRightArrow not and and { ifill } if
brushNone not { istroke } if
0 0 1 1 leftarrow
n 2 sub dup n 1 sub dup rightarrow
} if
end
} dup 0 4 dict put def

/Circ {
newpath
0 360 arc
patternNone not { ifill } if
brushNone not { istroke } if
} def

/CBSpl {
0 begin
dup 2 gt {
storexyn
newpath
n 1 sub dup 0 0 1 1 2 2 true subspline
1 1 n 3 sub {
/i exch def
i 1 sub dup i dup i 1 add dup i 2 add dup false subspline
} for
n 3 sub dup n 2 sub dup n 1 sub dup 0 0 false subspline
n 2 sub dup n 1 sub dup 0 0 1 1 false subspline
patternNone not { ifill } if
brushNone not { istroke } if
} {
Poly
} ifelse
end
} dup 0 4 dict put def

/Elli {
0 begin
newpath
4 2 roll
translate
scale
0 0 1 0 360 arc
patternNone not { ifill } if
brushNone not { istroke } if
end
} dup 0 1 dict put def

/Line {
0 begin
2 storexyn
newpath
x 0 get y 0 get moveto
x 1 get y 1 get lineto
brushNone not { istroke } if
0 0 1 1 leftarrow
0 0 1 1 rightarrow
end
} dup 0 4 dict put def

/MLine {
0 begin
storexyn
newpath
n 1 gt {
x 0 get y 0 get moveto
1 1 n 1 sub {
/i exch def
x i get y i get lineto
} for
patternNone not brushLeftArrow not brushRightArrow not and and { ifill } if
brushNone not { istroke } if
0 0 1 1 leftarrow
n 2 sub dup n 1 sub dup rightarrow
} if
end
} dup 0 4 dict put def

/Poly {
3 1 roll
newpath
moveto
-1 add
{ lineto } repeat
closepath
patternNone not { ifill } if
brushNone not { istroke } if
} def

/Rect {
0 begin
/t exch def
/r exch def
/b exch def
/l exch def
newpath
l b moveto
l t lineto
r t lineto
r b lineto
closepath
patternNone not { ifill } if
brushNone not { istroke } if
end
} dup 0 4 dict put def

/Text {
ishow
} def

/idef {
dup where { pop pop pop } { exch def } ifelse
} def

/ifill {
0 begin
gsave
patternGrayLevel -1 ne {
fgred bgred fgred sub patternGrayLevel mul add
fggreen bggreen fggreen sub patternGrayLevel mul add
fgblue bgblue fgblue sub patternGrayLevel mul add setrgbcolor
eofill
} {
eoclip
originalCTM setmatrix
pathbbox /t exch def /r exch def /b exch def /l exch def
/w r l sub ceiling cvi def
/h t b sub ceiling cvi def
/imageByteWidth w 8 div ceiling cvi def
/imageHeight h def
bgred bggreen bgblue setrgbcolor
eofill
fgred fggreen fgblue setrgbcolor
w 0 gt h 0 gt and {
l b translate w h scale
w h true [w 0 0 h neg 0 h] { patternproc } imagemask
} if
} ifelse
grestore
end
} dup 0 8 dict put def

/istroke {
gsave
brushDashOffset -1 eq {
[] 0 setdash
1 setgray
} {
brushDashArray brushDashOffset setdash
fgred fggreen fgblue setrgbcolor
} ifelse
brushWidth setlinewidth
originalCTM setmatrix
stroke
grestore
} def

/ishow {
0 begin
gsave
fgred fggreen fgblue setrgbcolor
/fontDict printFont printSize scalefont dup setfont def
/descender fontDict begin 0 [FontBBox] 1 get FontMatrix end
transform exch pop def
/vertoffset 1 printSize sub descender sub def {
0 vertoffset moveto show
/vertoffset vertoffset printSize sub def
} forall
grestore
end
} dup 0 3 dict put def
/patternproc {
0 begin
/patternByteLength patternString length def
/patternHeight patternByteLength 8 mul sqrt cvi def
/patternWidth patternHeight def
/patternByteWidth patternWidth 8 idiv def
/imageByteMaxLength imageByteWidth imageHeight mul
stringLimit patternByteWidth sub min def
/imageMaxHeight imageByteMaxLength imageByteWidth idiv patternHeight idiv
patternHeight mul patternHeight max def
/imageHeight imageHeight imageMaxHeight sub store
/imageString imageByteWidth imageMaxHeight mul patternByteWidth add string def
0 1 imageMaxHeight 1 sub {
/y exch def
/patternRow y patternByteWidth mul patternByteLength mod def
/patternRowString patternString patternRow patternByteWidth getinterval def
/imageRow y imageByteWidth mul def
0 patternByteWidth imageByteWidth 1 sub {
/x exch def
imageString imageRow x add patternRowString putinterval
} for
} for
imageString
end
} dup 0 12 dict put def

/min {
dup 3 2 roll dup 4 3 roll lt { exch } if pop
} def

/max {
dup 3 2 roll dup 4 3 roll gt { exch } if pop
} def

/midpoint {
0 begin
/y1 exch def
/x1 exch def
/y0 exch def
/x0 exch def
x0 x1 add 2 div
y0 y1 add 2 div
end
} dup 0 4 dict put def

/thirdpoint {
0 begin
/y1 exch def
/x1 exch def
/y0 exch def
/x0 exch def
x0 2 mul x1 add 3 div
y0 2 mul y1 add 3 div
end
} dup 0 4 dict put def

/subspline {
0 begin
/movetoNeeded exch def
y exch get /y3 exch def
x exch get /x3 exch def
y exch get /y2 exch def
x exch get /x2 exch def
y exch get /y1 exch def
x exch get /x1 exch def
y exch get /y0 exch def
x exch get /x0 exch def
x1 y1 x2 y2 thirdpoint
/p1y exch def
/p1x exch def
x2 y2 x1 y1 thirdpoint
/p2y exch def
/p2x exch def
x1 y1 x0 y0 thirdpoint
p1x p1y midpoint
/p0y exch def
/p0x exch def
x2 y2 x3 y3 thirdpoint
p2x p2y midpoint
/p3y exch def
/p3x exch def
movetoNeeded { p0x p0y moveto } if
p1x p1y p2x p2y p3x p3y curveto
end
} dup 0 17 dict put def

/storexyn {
/n exch def
/y n array def
/x n array def
n 1 sub -1 0 {
/i exch def
y i 3 2 roll put
x i 3 2 roll put
} for
} def

/SSten {
fgred fggreen fgblue setrgbcolor
dup true exch 1 0 0 -1 0 6 -1 roll matrix astore
} def

/FSten {
dup 3 -1 roll dup 4 1 roll exch
newpath
0 0 moveto
dup 0 exch lineto
exch dup 3 1 roll exch lineto
0 lineto
closepath
bgred bggreen bgblue setrgbcolor
eofill
SSten
} def

/Rast {
exch dup 3 1 roll 1 0 0 -1 0 6 -1 roll matrix astore
} def

%%EndProlog

%%BeginIdrawPrologue
/arrowhead {
0 begin
transform originalCTM itransform
/taily exch def
/tailx exch def
transform originalCTM itransform
/tipy exch def
/tipx exch def
/dy tipy taily sub def
/dx tipx tailx sub def
/angle dx 0 ne dy 0 ne or { dy dx atan } { 90 } ifelse def
gsave
originalCTM setmatrix
tipx tipy translate
angle rotate
newpath
arrowHeight neg arrowWidth 2 div moveto
0 0 lineto
arrowHeight neg arrowWidth 2 div neg lineto
patternNone not {
originalCTM setmatrix
/padtip arrowHeight 2 exp 0.25 arrowWidth 2 exp mul add sqrt brushWidth mul
arrowWidth div def
/padtail brushWidth 2 div def
tipx tipy translate
angle rotate
padtip 0 translate
arrowHeight padtip add padtail add arrowHeight div dup scale
arrowheadpath
ifill
} if
brushNone not {
originalCTM setmatrix
tipx tipy translate
angle rotate
arrowheadpath
istroke
} if
grestore
end
} dup 0 9 dict put def

/arrowheadpath {
newpath
arrowHeight neg arrowWidth 2 div moveto
0 0 lineto
arrowHeight neg arrowWidth 2 div neg lineto
} def

/leftarrow {
0 begin
y exch get /taily exch def
x exch get /tailx exch def
y exch get /tipy exch def
x exch get /tipx exch def
brushLeftArrow { tipx tipy tailx taily arrowhead } if
end
} dup 0 4 dict put def

/rightarrow {
0 begin
y exch get /tipy exch def
x exch get /tipx exch def
y exch get /taily exch def
x exch get /tailx exch def
brushRightArrow { tipx tipy tailx taily arrowhead } if
end
} dup 0 4 dict put def

%%EndIdrawPrologue

%I Idraw 10 Grid 8 8

%%Page: 1 1

Begin
%I b u
%I cfg u
%I cbg u
%I f u
%I p u
%I t
[ 0.719734 0 0 0.719734 0 0 ] concat
/originalCTM matrix currentmatrix def

Begin %I Rect
%I b 65535
2 0 0 [] 0 SetB
%I cfg Black
0 0 0 SetCFg
%I cbg White
1 1 1 SetCBg
none SetP %I p n
%I t
[ 0.707107 0.707107 -0.707107 0.707107 671.26 222.652 ] concat
%I
76 290 260 471 Rect
End

Begin %I Rect
%I b 65535
2 0 0 [] 0 SetB
%I cfg Black
0 0 0 SetCFg
%I cbg White
1 1 1 SetCBg
none SetP %I p n
%I t
[ 0.707107 0.707107 -0.707107 0.707107 415.174 220.859 ] concat
%I
74 291 259 471 Rect
End

Begin %I Line
%I b 65535
2 0 0 [] 0 SetB
%I cfg Black
0 0 0 SetCFg
%I cbg White
1 1 1 SetCBg
none SetP %I p n
%I t
[ 1 -0 -0 1 97 227 ] concat
%I
166 252 424 252 Line
%I 1
End

Begin %I Text
%I cfg Black
0 0 0 SetCFg
%I f *-helvetica-medium-r-*-140-*
Helvetica 14 SetF
%I t
[ 0.707107 0.707107 -0.707107 0.707107 299.912 548.555 ] concat
%I
[
(HORIZON)
] Text
End

Begin %I Text
%I cfg Black
0 0 0 SetCFg
%I f *-helvetica-medium-r-*-140-*
Helvetica 14 SetF
%I t
[ 1 0 0 1 304 724 ] concat
%I
[
(DESITTER REGION)
] Text
End

Begin %I Text
%I cfg Black
0 0 0 SetCFg
%I f *-helvetica-medium-r-*-140-*
Helvetica 14 SetF
%I t
[ 1 0 0 1 370 692 ] concat
%I
[
(s<0)
] Text
End

Begin %I Text
%I cfg Black
0 0 0 SetCFg
%I f *-helvetica-medium-r-*-140-*
Helvetica 14 SetF
%I t
[ 1 0 0 1 372 535 ] concat
%I
[
(s<0)
] Text
End

Begin %I Text
%I cfg Black
0 0 0 SetCFg
%I f *-helvetica-medium-r-*-140-*
Helvetica 14 SetF
%I t
[ 1 0 0 1 505 629 ] concat
%I
[
(s>0)
] Text
End

Begin %I Text
%I cfg Black
0 0 0 SetCFg
%I f *-helvetica-medium-r-*-140-*
Helvetica 14 SetF
%I t
[ 1 0 0 1 218 627 ] concat
%I
[
(s>0)
] Text
End

Begin %I Text
%I cfg Black
0 0 0 SetCFg
%I f *-helvetica-medium-r-*-140-*
Helvetica 14 SetF
%I t
[ 1 0 0 1 139 362 ] concat
%I
[
(Figure 2. Penrose diagram for a quantum black hole in equilibrium )
()
(with a radiation bath. The singularity present in the classical Kruskal)
()
(diagram is replaced by an asymptotically deSitter region.)
] Text
End

Begin %I Line
%I b 65535
2 0 0 [] 0 SetB
%I cfg Black
0 0 0 SetCFg
%I cbg White
1 1 1 SetCBg
none SetP %I p n
%I t
[ 1 -0 -0 1 96 225 ] concat
%I
169 513 425 513 Line
%I 1
End

End %I eop

showpage

%%Trailer

end